# Discrimination-Based Double Auction for Maximizing Social Welfare in the Electricity and Heating Market Considering Privacy Preservation

Lu Wang, *Graduate Student Member, IEEE*, Wei Gu, *Senior Member, IEEE*, Shuai Lu, *Member, IEEE*, Haifeng Qiu, *Member, IEEE*, and Zhi Wu, *Member, IEEE*

*Abstract*—This paper proposes a doubled-sided auction mechanism with price discrimination for social welfare (SW) maximization in the electricity and heating market. In this mechanism, energy service providers (ESPs) submit offers and load aggregators (LAs) submit bids to an energy trading center (ETC) to maximize their utility; in turn, the selfless ETC as an auctioneer leverages discriminatory price weights to regulate the behaviors of ESPs and LAs, which combines the individual benefits of each stakeholder with the overall social welfare to achieve the global optimum. Nash games are employed to describe the interactions between players with the same market role. Theoretically, we first prove the existence and uniqueness of the Nash equilibrium; then, considering the requirement of game players to preserve privacy, a distributed algorithm based on the alternating direction method of multipliers is developed to implement distributed bidding and analytical target cascading algorithm is applied to reach the balance of demand and supply. We validated the proposed mechanism using case studies on a city-level distribution system. The results indicated that the achieved SW improved by 4%-15% compared with other mechanisms, and also verified the effectiveness of the distributed algorithm.

## NOMENCLATURE

### A. Abbreviation

| | |
|---|---|
| LA | Load aggregator |
| ESP | Energy service provider |
| ETC | Energy trading center |
| MES | Multi-energy system |
| SW | Social welfare |
| DADP | Double-sided auction mechanism with discriminatory pricing |
| KEL | Kelly mechanism |
| NE | Nash equilibrium |
| ADBA | ADMM-based distributed bidding algorithm |

This work was supported by the Smart Grid Joint Fund of National Science Foundation of China & State Grid Corporation of China (U1866208); supported by the China scholarship Council. (*Corresponding author: Wei Gu.*)

L. Wang, W. Gu, S. Lu, and Z. Wu are with the School of Electrical Engineering, Southeast University, Nanjing 210096, China. (luw0804@gmail.com, wgu@seu.edu.cn, lushuai1004@outlook.com, zwu@seu.edu.cn). H. Qiu was with the School of Electrical Engineering, Southeast University, Nanjing 210096, China, and is with the School of Electrical and Electronic Engineering, Nanyang Technological University, Singapore 639798. (895416086@qq.com).

### B. Sets and Indices

| | |
|---|---|
| $\Omega_i$ | Set of indices of LAs, indexed by $i$ |
| $\Omega_j$ | Set of indices of ESPs, indexed by $j$ |
| $\Omega_i^{LA}$ | The feasible strategy set of LA $i$ |
| $\Omega_j^{ESP}$ | The feasible strategy set of ESP $j$ |
| $-i, -j$ | The rest of LA $-i \in \Omega_{i,-i\neq i}$ except for LA $i \in \Omega_i$ / the rest of ESP $-j \in \Omega_{j,-j\neq j}$ except for ESP $j \in \Omega_j$ |
| $I, J$ | Total number of LAs/ ESPs |
| $D, S$ | Constraints for LAs/ ESPs |
| $\boldsymbol{a}, \boldsymbol{b}$ | Set of quotes from ESPs/ LAs |
| $v_i$ | The value function of LA $i$ ($) |
| $c_j$ | The cost function of ESP $j$ ($) |
| $u_i, u_j$ | Payoff function of LA $i$/ESP $j$ ($) |
| $\aleph_D^E, \aleph_S^E$ | The known information set of ETC in demand side / supply side |
| $\aleph_i^D$ | The known information set of LA $i$ in demand side |
| $\aleph_j^S$ | The known information set of ESP $j$ in supply side |

### C. Parameters

| | |
|---|---|
| $d_i^{min}, s_j^{max}$ | The minimum demand of LA $i$/ maximum supply of ESP $j$ (MWh) |
| $T_{max}^{in}, T_{min}^{in}$ | Adjustable range of the indoor temperature (°C) |
| $\alpha, \beta$ | Primary and quadratic coefficients of the value function |
| $m, n$ | Quadratic and primary coefficients of the cost function |
| $\chi, \gamma$ | Multipliers of Lagrangian penalty function |
| $R$ | Thermal resistance of the house shell (°C/MW) |
| $C$ | Heat capacity of the air (MWh/°C) |
| $\epsilon$ | Convergence tolerance |
| $\Delta t$ | Time interval (h) |

### D. Variables

| | |
|---|---|
| $b_i, a_j$ | Heat (electricity) quote of LA $i$/ESP $j$ ($) |
| $b_{-i}, a_{-j}$ | Heat (electricity) quote from the rest of LA $-i \in \Omega_{i,-i\neq i}$ except LA $i \in \Omega_i$ / the rest of ESP $-j \in \Omega_{j,-j\neq j}$ except ESP $j \in \Omega_j$ ($) |
| $p_i, q_j$ | Price weight of heat (electricity) sent by ETC to LA $i$ / ESP $j$ |
| $p_{-i}, q_{-j}$ | Price weight of heat (electricity) sent by ETC to LA $-i \in \Omega_{i,-i\neq i}$ except for LA $i \in \Omega_i$/ ESP $-j \in \Omega_{j,-j\neq j}$ except for ESP $j \in \Omega_j$ |
| $p_i^*, q_j^*$ | Price weight of heat (electricity) sent by ETC to LA $i$ / ESP $j$ of NE points |
| $\mu(b), \omega(a)$ | Estimation of heat (electricity) clearing price per unit of LA/ ESP by ETC ($/MWh) |
| $d_i, s_j$ | The heat (electricity) demand of LA $i$/ supply of ESP $j$ (MWh) |
| $d_{-i}, s_{-j}$ | The heat (electricity) demand/supply of LA $-i \in \Omega_{i,-i\neq i}$ except for LA $i \in \Omega_i$/ ESP $-j \in \Omega_{j,-j\neq j}$ except for ESP $j \in \Omega_j$ (MWh) |
| $d_i^*, s_j^*$ | The heat (electricity) demand/supply of LA $i$ / ESP $j$ at Nash equilibrium points |
| $T_t^{in}, T_t^{out}$ | Indoor and outdoor temperatures at period $t$ (°C) |
| $P_t^{min}, P_t^{max}$ | Adjustable range of the heat demand |
| $z, x$ | Auxiliary vector |
| $\rho$ | Penalty parameters |

## I. INTRODUCTION

To overcome the challenges of widespread energy poverty and the intensifying climate disruption, the multi-energy system (MES) [1], which harnesses multiple energy sources, has emerged as a new energy system that contributes a great deal to smart and green cities. Studies on the MES have focused on operation [2], planning [3], economic dispatching [4], and trading [5]. The development of the MES has precipitated the advent of the 'energy service provider (ESP)', which combines the multiple functions of the production, conversion and transaction of different energy carriers. The proliferation of ESPs motivates a promising paradigm of energy distribution, for which a fair multi-energy coupling trading system and an efficient auction mechanism design are imminent.

The auction mechanism has been proved in practice to allocate distributed resources effectively [6], which can be roughly categorized into single-sided auctions and double-sided auctions based on the features of the bidders. A market environment where only one type of player participates in the bidding can be categorized as a single-sided auction, such as in the traditional centralized power market where only buyers bid [7]. For the single-sided auction, considering the time-shifting load, a price-setting strategy for the multi-time scale market was designed for price makers in [8]. Different from [8], a joint bid for flexible loads and photovoltaic energy was placed on the local electricity market in [9]. Coincidentally, for the multi-energy market, an independent auction mechanism was designed in [5] for two-time-scale markets to maximize social welfare (SW). Moreover, a market layer was added between the supply and demand layers in [10] to study the distributed electric-heating market more closely. The model structure proposed in [11] is similar to [10], which innovatively used mathematical programming with equilibrium constraints to characterize the power-heat transaction while using the Stackelberg game to characterize the one-sided auction mechanism. However, it is unfair to have a party that has a resource advantage that masters scarce resources.

In contrast to single-sided auctions, double-sided auctions have more than one type of participant, with both buyers and sellers participating in the bidding. The double-sided auction is more realistic than the single-sided auction [12], and it has been widely adopted in the European day-ahead market and intraday market [13]. Azizi et al. [14] proposed a transactive market mechanism with loss allocation for the electric distribution network, which has solved the cross-subsidization problem in a centralized manner. To face the complex market environment, an optimized pricing mechanism was implemented in [15] that included two types of trading networks: a local trading network and an external large grid trading network. For the integrated energy market, optimization models for multi-energy companies and multiple users were designed in [16] for the power-heat market and power-gas market to increase the synergy between multiple energy sources. Specifically, hierarchical optimization [17] and game theory are widely used in auctions to characterize stakeholder interactions. The system-level interaction in the power-heat market developed in [18] was captured by a Nash game. A two-layer demand-response market was extended in [19], which included single-sided auctions and double-sided auctions, using a three-step algorithm to model the interactions among stakeholders. A generalized Nash game was investigated in [20] to characterize the double-sided auction among prosumers without role restrictions. However, the selfish bidding behavior of stakeholders can make the market optimal in terms of social welfare only if the market is perfectly competitive [21]. In summary, most existing studies are based on the assumption of perfect competition, which is inconsistent with the reality of energy markets.

On the other hand, distributed algorithms have been widely applied in the field of energy trading, such as the distributed consensus method [22], alternating direction method of multipliers (ADMM) and analytical target cascading (ATC). The ADMM algorithm was explored in [23] to solve the distributed multi-class energy management problem for scalability and privacy consideration. An improved ADMM algorithm was proposed in [24] to achieve the energy management of a multi-energy coupling network with a completely distributed method. The ATC algorithm was developed in [25] to enable multi-stakeholder cooperation to provide multiple energy sources without a coordinator. A market-oriented program can be interpreted as distributed optimization [26], in which every trader should be regarded as a rational individual. As a result, players are reluctant to share their private information during the bidding process.

While emerging communications and distributed algorithms are ripe for use in marketplace bidding, the issue of personal privacy protection has not been adequately addressed. Moreover, with the deregulation of the energy market, different types of energy systems are increasingly being connected. However, each energy system is under different management and has different interests to pursuit, making it more urgent to address privacy protection in transactions. In this circumstance, several issues of energy trading in MES have not been fully addressed, as follows:

1) In the energy trading process of MES, although the integrated demand response has been widely considered [27], the demand side is usually regarded as a price taker, which inhibits their participation in trading.

2) A game theory similar to the Starkelberg game has been used to formulate the competition among stakeholders in the MES market [13]. However, most efforts either focus on the benefits of the players [12] or ignore individual interests to maximize the overall SW [28]. Due to the limited number of players, the practical MES market is imperfectly competition and the selfish pursuit of self-interest by all parties causes SW losses [29].

3) The auction mechanisms which encompass individual rationality and SW maximization have been applied in the energy market, such as Vickrey-Clark-Groves (VCG) [5] and Trade Reduction [30], while the budget balance cannot be guaranteed and the private information of players is not well protected [31].

To overcome the challenges mentioned above, in this paper, a double-sided auction mechanism with discriminatory pricing (DADP) is designed for power-heating trading to improve SW. To protect the privacy information, an ADMM-based distributed bidding is proposed to coordinate various players. Furthermore, the ATC algorithm is applied to strike a balance between supply and demand. Note that in this article, we use "privacy-preserving" for all transaction participants: ESPs and LAs, to keep their private information from being disclosed. The definition of privacy-preserving is: *no player can directly obtain or indirectly deduce any private information, including cost function, value function, exact demand and exact supply of other players from the information they have.* Specific detailed definitions are given in Section IV. The information is classified as *fully public*, *private*, and *semi-public* depending on the level of sensitivity. For each player ESP on the supply-side, the individual cost functions and supply price weights are given by the ETC are private information, the offers are public information, and the estimated total demand is known to all ESPs as semi-public information. For any player LA on the demand-side, their individual value function and demand price weights set by the ETC are private information, the bids are public information, and the estimated total supply is known to all LAs as semi-public information.

The main contributions are as follows:

1) A novel one-dimensional bidding strategy is proposed to allow both suppliers and consumers to trade energy in the market as a price setter. Unlike the existing bidding strategies that provide both price and quantity, this method only provides bidding price parameters to protect the private information of stakeholders.

2) A novel market-oriented DADP is designed for heterogeneous energy trading, which combines the selfish behavior of stakeholders with social welfare to mitigate the loss of SW. Consumers with high values and suppliers with low costs are incentivized to expand market share according to the price weights set by the ETC.

3) The existence and uniqueness of the Nash equilibrium (NE) of the demand-side and supply-side are proved. A distributed algorithm based on ADMM is exploited to implement distributed bidding with privacy considerations, whereas the ATC algorithm is applied to achieve a balance between supply and demand.

The remainder of this paper is organized as follows: In Section II, the system model is presented. In Section III, the game model of the double-sided auction is established, and the existence and uniqueness of the NE are proved. A distributed algorithm is proposed in Section IV and case studies are presented in Section V. In Section VI, the paper is concluded.

## II. SYSTEM MODEL

### A. Assumptions

A DADP mechanism in the power-heat market is envisioned. The relevant assumptions of this paper are as follows:

1) LAs and ESPs are rational and avoid disclosing private information in transactions. As an auctioneer, the ETC is a selfless institution that balances supply and demand.

2) Considering the reality of multi-energy transactions in MES, the number of various types of participants in the transaction is limited. Information barriers exist in the market, and transaction information is not fully disclosed. Thus, the whole process of market competition is considered as imperfect competition rather than perfect competition which does not exist in practice.

3) The electricity and heat markets are cleared independently, and integrated energy demand response is considered. The electrical load of LAs consists of the fixed load and the flexible load, and then heating load takes into account the thermal comfort of the consumers.

4) The value function of each LA is continuous, strictly monotonically increasing in the domain of definition of the independent variable $d_i$, and is a concave function in the interval $[0, +\infty)$. The cost function of each ESP is continuous, strictly monotonously increasing, satisfies $c(s_j) = 0$, when $s_j \leq 0$, and is a convex function in the interval $[0, +\infty)$.

5) When ESPs participate in the market offer, the number of ESPs satisfies $J > 2$, and the offer of all ESPs participating in the market satisfies $a_j \neq 0$.

6) This paper focuses on the design of energy market mechanisms, where electricity and heat are considered to be traded without distinction at the mechanism design level. In contrast, electricity and heat are considered to be differentiated at the demand level based on the different ways of participating in demand response.

### B. Auction Round Implementation

Consider a multi-energy market with three agents: the auctioneer, a demand-side set of LAs $\Omega_i = \{1, ..., I\}$, and a supply-side set of ESPs $\Omega_j = \{1, ..., J\}$. An LA aggregates residential units, utilities and other facilities containing distributed energy in the region to form a consortium that can participate in market bidding. LAs can perform a thermal demand response inside the buildings and a distributed generation unit outside the building to participate in electricity demand response. The ESP contains, for example, various distributed resources, energy storage, and energy conversion equipment, which can provide power and heat. The power and heat market are independent of each other, which conforms to real-world market operations. In this case, a DADP mechanism is designed for market clearing with the following process. The auction process is shown in Fig. 1.

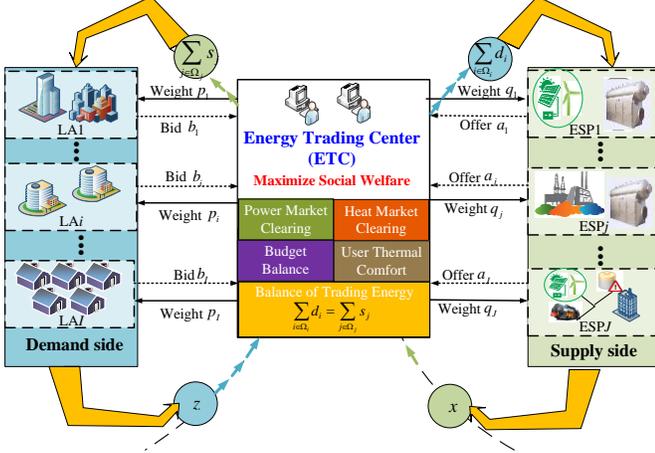

Fig. 1. Double-sided auction process for heat and power.

The ETC first advertises the initial estimated total supply $z \approx \sum_{j \in \Omega_j} s_j$ and price weight $p_i$ to the demand side as a trusted party auctioneer. The LAs conduct strategic bidding $b_i \in \boldsymbol{b} = \{b_1, \dots, b_I\}$ based on the obtained market information to maximize energy consumption value, where the offer is a monetized measure of their willingness to purchase energy. Specifically, value functions are introduced to express energy utilization in monetary terms,

$$v_i(d_i) = \alpha_i d_i - \beta_i d_i^2 \tag{1}$$

which are quadratic functions [32] that satisfy assumption 4. Subsequently, all bids $b$ are submitted to the ETC. The bids of LAs are indexed by $b \in [b_i]$, where $b_i$ is LA $i$'s bid price for the demand. Then, the ETC sends the estimated total demand $x \approx \sum_{i \in \Omega_i} d_i$ and price weight $q_j$ to the supply side. Following the market information, ESPs offer $a_j \in \boldsymbol{a} = \{a_1, \dots, a_J\}$ based on the cost function, indicating the extent to which the ESP is unwilling to provide energy, measured in dollars, which is then returned to the ETC. Explicitly, a quadratic function is used to represent the cost of ESP as follows [33],

$$c_j(s_j) = \begin{cases} m_j s_j^2 + n_j s_j, & s_j > 0 \\ 0, & s_j \leq 0 \end{cases} \tag{2}$$

which satisfied assumption 4.

The ETC determines $\omega(\boldsymbol{a})$ and $\mu(\boldsymbol{b})$, which are the shadow prices of trading energy in economic theory, while the market weights $p_i$, $q_j$ are updated independently, depending on the submitted bid price $\boldsymbol{b}$ and offer price $\boldsymbol{a}$. Finally, it adapts to the NE through multiple iterations. ETC manipulates the amount of energy allocated to each entity and the fees charged/paid to the entity using a discriminatory price weight.

The goal of the DADP mechanism is to combine the interests of players with SW without the private information of players. The whole process of implementing the DADP mechanism can be formulated as followed:

$$\max_{d,s} \sum_{i \in \Omega_i} v_i(d_i) - \sum_{j \in \Omega_j} c_j(s_j) \tag{3a}$$

$$s.t. \sum_{i \in \Omega_i} d_i = \sum_{j \in \Omega_j} s_j \tag{3b}$$

$$d_i^{\min} \leq d_i, s_j \leq s_j^{\max} \tag{3c}$$

$$\sum_{j \in \Omega_j} q_j a_j(d_j) \leq \sum_{i \in \Omega_i} p_i b_i(s_i) \tag{3d}$$

$$d_i \in D, s_i \in S, \tag{3e}$$

constraint (3b) represents the energy balance, (3c) represents supply and demand constraints, and the bid $a_j$ of LA in (3d) is a function of demand $d_j$ and the offer $b_i$ of ESP is a function of supply $s_i$. This constraint ensures that the deal is concluded, from an economic point of view, to achieve budget balance in the auction [34]. And the constraint (3e) denotes some relevant local constraints on the demand $d_j$ and the supply $s_i$.

The problem of maximizing SW (3) is decomposed into subproblems, demand-side utility maximization corresponding to the value function and supply-side utility maximization corresponding to the cost function, which can be formulated as demand-side auction and supply-side auction, respectively.

### C. Supply Side Auction Mechanism

Maximizing SW on the supply side can be characterized by minimizing the cost of ESPs payments, the objective function (4a) is defined in (2):

$$\min_s \sum_{j \in \Omega_j} c_j(s_j) \tag{4a}$$

$$s.t. \quad s \in S \triangleq \left\{ s \in \mathbb{R}^J \,\bigg|\, \sum_{j \in \Omega_j} s_j = \sum_{i \in \Omega_i} d_i, s_j \leq s_j^{\max} \right\}. \tag{4b}$$

The solution to (4) is to maximize the supply-side utility. The offer $a_j$ submitted by ESP $j$ is related to the supply function [35]:

$$S(\omega(a_j); a_j) = \sum_{i \in \Omega_i} d_i - a_j / \omega(a_j), \tag{5}$$

where $S(\omega(a_j); a_j)$ denotes the supply function, the quotation $a_j$ is measured in US dollars which reflects the ability of ESP to supply energy in the market, and $\omega(a_j)$ is the estimation of clearing price per unit of energy determined by the ETC based on offer $a_j$.

According to supply function (5) and balance of supply and demand $\sum_{j \in \Omega_j} S(\omega(a_j); a_j) = \sum_{i \in \Omega_i} d_i$, when all bids $a$ are given, from (5) we have $\omega(\boldsymbol{a}) = \sum_{j \in \Omega_j} a_j / (J-1) \sum_{i \in \Omega_i} d_i$. The vector $\boldsymbol{a}$ is the offer of all ESPs, that is $\boldsymbol{a} = (a_j, a_{-j})$, and the notation $a_{-j}$ is used to indicate the offer of ESPs other than ESP $j$, $a_{-j} = (a_1, \dots, a_{j-1}, a_{j+1}, \dots, a_J)$. Given $a_{-j}$, the energy supplied by each ESP is derived as followed:

$$S(\omega(\boldsymbol{a}); a_j) = \sum_{i \in \Omega_i} d_i - \frac{a_j}{\sum_{j \in \Omega_j} a_j}(J-1) \sum_{i \in \Omega_i} d_i, \tag{6}$$

where $J$ is the total number of ESPs, the ESP's offer reflects its energy supply cost. Additionally, the actual unit clearing price of ESP $j$ is $q_j \omega(\boldsymbol{a})$, which is the product of price weight and clearing price valuation per unit of energy. The utility function for ESP $j$ is as follows:

$$u_j(\boldsymbol{a}) = q_j \cdot \omega(\boldsymbol{a}) \cdot S(\omega(\boldsymbol{a}); a_j) - c_j(S(\omega(\boldsymbol{a}); a_j)), \quad (7)$$

then the expression for $\omega(\boldsymbol{a})$ and eq. (6) are brought into (7), we can get the following equation,

$$u_j(\boldsymbol{a}) = q_j \cdot \left( \frac{\sum_{j \in \Omega_j} a_j}{J-1} - a_j \right) - c_j \left( \sum_{i \in \Omega_i} d_i - \frac{a_j(J-1)}{\sum_{j \in \Omega_j} a_j} \sum_{i \in \Omega_i} d_i \right), a_{-j} \neq 0, \quad (8)$$

where the first part represents the revenue from the energy supply, and the second part represents the energy cost of ESP $j$.

***Remark 1:*** For the supply-side stakeholders, the private information of ESP $j$ is the cost function $c_j$ whose disclosure would put ESP $j$ at a competitive disadvantage in the market bidding process. The accurate supply quantity $s_j$ is only known by ESP $j$, and that can only be estimated by ETC, while the demand-side LAs only know the estimated sum of the supply quantities $z \approx \sum_{j \in \Omega_j} s_j$ and do not know the exact supply quantity of each ESP. In addition, the estimation of the clearing price per unit of energy $\omega(a_j)$ determined by the ETC as *semi-public information*, which is only available between ETC and ESP $j$. The offer $\boldsymbol{a} = \{a_j, a_{-j}\}, \forall j, -j \in \Omega_{j,-j \neq j}$ of each ESP is *fully public information*.

*D. Demand Side Auction Mechanism*

Maximizing SW on the demand-side maximizes the cumulative value obtained by the LAs, the objective function is defined as (1):

$$\max_{d} \sum_{i \in \Omega_i} v_i(d_i) \quad (9a)$$

$$s.t. \quad d \in D \triangleq \left\{ d \in \mathbb{R}^I \middle| \sum_{i \in \Omega_i} d_i = \sum_{j \in \Omega_j} s_j, d_i^{\min} \leq d_i \right\}. \quad (9b)$$

constraint (9b) ensures that users' demands are met.

The LA submits bid $b_i$, and the demand function is interpreted as the energy demand of LA, when the unit market-clearing price is $\mu$ [35]:

$$D(\mu(b_i); b_i) = b_i / \mu(b_i), \quad (10)$$

where $D(\mu(b_i); b_i)$ denotes the demand function, $b_i$ is measured in US dollars which implies the ability of LA to consume energy in the market, and $\mu(b_i)$ is the estimation of clearing price per unit of energy determined by the ETC based on bid $b_i$.

When all bids $\boldsymbol{b}$ are given, where $\boldsymbol{b} = (b_i, b_{-i})$, the notation $b_{-i}$ is used to indicate the offer of ESPs other than ESP $i$, $b_{-i} = (b_1, \ldots, b_{i-1}, b_{i+1}, \ldots, b_I)$, and from (10) we have $\mu(\boldsymbol{b}) = \sum_{i \in \Omega_i} b_i / \sum_{i \in \Omega_i} D$. Considering the balance of energy $\sum_{i \in \Omega_i} D(\mu(b_i); b_i) = \sum_{j \in \Omega_j} s_j$, the energy obtained by each LA is

$$D(\mu(\boldsymbol{b}); b_i)) = \frac{b_i}{\sum_{i \in \Omega_i} b_i} \sum_{j \in \Omega_j} s_j. \quad (11)$$

The LA reflects its energy purchasing strategy through the bid, meanwhile, the actual unit clearing price is $p_i \mu(\boldsymbol{b})$, which is the product of the price weight of LA $i$ and the valuation of the clearing price. The utility function for LA $i$ is as follows:

$$u_i(\boldsymbol{b}) = v_i \left( \frac{b_i}{\sum_{i \in \Omega_i} b_i} \sum_{j \in \Omega_j} s_j \right) - p_i b_i, b_{-i} \neq 0, \quad (12)$$

where the first part represents the energy value of LA $i$ and the second part denotes the payment of the LA to purchase energy.

***Remark 2:*** For the demand-side stakeholders, the value functions $v_i$ is *privacy information*, which is not willing to be shared from the bidding perspective. In contrast, the demand $d_i$ of each LA $i$ is *semi-public information*, which can be estimated by ETC. Meanwhile, the supply-side ESPs only have access to the estimated sum demand $x \approx \sum_{i \in \Omega_i} d_i$ and do not know the definite demand of each LA. And $\mu(b_i)$ is the estimation of clearing price per unit of energy, which is the semi-public only accessed by ETC and LA $i$. Besides, the *fully public information* is the bid price $b_i$ of each LA.

*E. User Thermal Comfort Model*

Currently, subscribers have two main ways to obtain heat, the first method is to purchase electricity from the electrical energy market and then convert it to heat by themselves, and the second approach is to purchase heat directly from the heating market, and these two ways are interchangeable.

The physical comfort of users depends on the heating supply, which is affected by the outdoor temperature and the insulation function of the building as follows [36]:

$$P^t = \left( \frac{T_{t+\Delta t}^{in} - T_t^{in} \cdot e^{-\Delta t/\tau}}{1 - e^{-\Delta t/\tau}} - T_t^{out} \right) 1/R, \quad (13)$$

where $\tau = R \cdot C$, $R$ is the thermal resistance of the house shell, $C$ is the heat capacity of the air, and $T_t^{in}$ and $T_t^{out}$ are the indoor and outdoor temperatures at period $t$, respectively.

Within the adjustable range of the indoor temperature $T_{max}^{in} \geq T_t^{in} \geq T_{min}^{in}$, to ensure the comfort of users at $t + \Delta t$, the adjustable range of the heat demand is as follows:

$$\begin{cases} P_{\min}^t = \left( \frac{T_{\min}^{in} - T_t^{in} \cdot e^{-\Delta t/\tau}}{1 - e^{-\Delta t/\tau}} - T_t^{out} \right) 1/R \\ P_{\max}^t = \left( \frac{T_{\max}^{in} - T_t^{in} \cdot e^{-\Delta t/\tau}}{1 - e^{-\Delta t/\tau}} - T_t^{out} \right) 1/R \end{cases}. \quad (14)$$

Consequently, the heating demand can participate in the market response, either through the choice of heat source, in detail electricity to heat or direct purchase of heat, or through the thermal regulation of the building's supply to participate in the demand response.

## III. PROBLEM FORMULATION

In the offering and bidding process, the ETC generates a game among players on the demand (supply) side by adjusting the weights $p$ and $q$. The demand-side game is denoted by $(I, p)$ and the supply-side game is denoted by $(J, q)$.

*A. Demand-Side Game Model*

From (11), the bids of the remaining LAs affect the utility function of LA $i$, and the game among the LAs is as follows:
1) *Players*: LAs in set $\Omega_i$.
2) *Strategies:* According to the price weight $p$ received, each

LA determines its energy purchasing plan and bids to maximize energy value. The strategy set of LA $i$ is

$$\Omega_i^{LA} = \{b(d) = (b_i(d_i), b_{-i}(d_{-i})) | \forall i, -i \in \Omega_{i,-i \neq i}, d \in D\} \quad (15)$$

where $b_i$ and $b_{-i}$ represent LA bids, $\boldsymbol{b} = \{b_i, b_{-i}\}$, $\boldsymbol{d} = \{d_i, d_{-i}\}, \forall i, -i \in \Omega_{i,-i \neq i}$, which only differentiated according to the quotes of different individuals $i, -i \in \Omega_{i,-i \neq i}$ of LA.

3) *Payoffs:* $u_i(b_i; b_{-i})$ is the payoff of LA $i$, as in (12).

**Definition 1:** When each LA satisfies $u_i(b_i; b_{-i}) \geq u_i(b_i'; b_{-i}), \forall b_i' \geq 0$, bid $\boldsymbol{b}$ is the NE of the demand-side game.

**Lemma 1:** For the demand-side game $(I, p)$, bid $\boldsymbol{b}$ is the NE if and only if the LA has

$$\frac{1}{p_i} v_i'(d_i)(1 - \frac{d_i}{\sum_{j \in \Omega_j} s_j}) \begin{cases} = \mu, d_i > 0 \\ \leq \mu, d_i = 0 \end{cases}. \quad (16)$$

***Proof of Lemma 1:***

1) *Necessity:* Assume that $\boldsymbol{b}$ is the NE and $b_{-i} \neq 0$ always holds for entity $i$; otherwise, assume that $b_{-i} = 0$ for a specific entity $i$ and its utility function is

$$u_i(b_i, b_{-i}) = \begin{cases} v_i(\frac{b_i}{\boldsymbol{b}} \sum_{j \in \Omega_j} s_j) - p_i b_i, b_i \neq 0 \\ v_i(0), b_i = 0 \end{cases}, \quad (17)$$

where $\boldsymbol{b}$ denotes the set of LA bids $\boldsymbol{b} = \{b_1, b_2, ..., b_I\}$, when a discontinuity exists at $b_{-i} = 0$, LA $i$ can always shape the bid to increase the utility, which contradicts the assumption that $\boldsymbol{b}$ is the NE. Therefore, $b_{-i} \neq 0$ is established for all LA $i$, and the utility function of the LA is:

$$u_i(b_i, b_{-i}) = v_i(\frac{b_i}{\sum_{i \in \Omega_i} b_i} \sum_{j \in \Omega_j} s_j) - p_i b_i, i \in \Omega_I, \quad (18)$$

which continuous at $b_i \in (0, +\infty)$, and its first partial derivative with respect to $b_i$ is

$$\frac{\partial}{\partial b_i} u_i(b_i, b_{-i}) = v_i'(\frac{b_i}{\sum_{i \in \Omega_i} b_i} \sum_{j \in \Omega_j} s_j) \frac{\sum_{i \in \Omega_i} b_i - b_i}{\sum_{i \in \Omega_i} b_i^2} \sum_{j \in \Omega_j} s_j - p_i$$

$$= \frac{p_i \sum_{j \in \Omega_j} s_j}{\sum_{i \in \Omega_i} b_i} [\frac{1}{p_i} v_i'(d_i)(1 - \frac{d_i}{\sum_{j \in \Omega_j} s_j}) - \mu] \quad (19)$$

The second-order partial derivative of $b_i$ is less than zero, so the utility function is strictly concave. The conclusion of the lemma follows from the extreme condition of the concave function.

2) *Sufficiency:* Prove that $d_i \in D$ holds for LA $i \in \Omega_i$; otherwise, assume that $d_i$ is taken to the maximum value $d_i = \sum_{j \in \Omega_j} s_j$, $d_{-i} = 0$ is true for all $-i \in \Omega_{-i \neq i}$. $\mu(b_i) = 0$ is known from the condition of $d_i$ in Lemma 1. For any $-i$, $\mu(b_{-i}) = v_{-i}'(0)/p_{-i} > 0$, which leads to a contradiction. Therefore, $b_{-i} \neq 0$ holds for LA $i \in \Omega_i$, and the utility function of the LA is

$$u_i(b_i, b_{-i}) = v_i(\frac{b_i}{\sum_{i \in \Omega_i} b_i} \sum_{j \in \Omega_j} s_j) - p_i b_i, i \in \Omega_I. \quad (20)$$

**Theorem 1:** On the demand side $(I, p)$, when $I > 1$ has a unique NE $\boldsymbol{b}$, for any NE $\boldsymbol{b}$, the vector $\boldsymbol{d}$ defined by $d_i = D(\mu(\boldsymbol{b}); b_i)$ is the only solution to (9), and is also the only solution of the variational inequality $VI\,(D, M)$, and $M\,(d, p)$ is defined as follows:

$$\hat{v}_i(d_i, p_i) \triangleq \int_0^{d_i} \frac{1}{p_i} v_i'(z)(1 - \frac{z}{\sum_{j \in \Omega_j} s_j}) dz \quad (21a)$$

$$M_i(d, p) \triangleq -\frac{\partial}{\partial d} \hat{v}_i(d_i, p_i) = -\frac{v_i'(d_i)}{p_i}(1 - d_i / \sum_{j \in \Omega_j} s_j). \quad (21b)$$

***Proof of Theorem 1:***

*Existence:* 1) $\Omega_i^{RU}$ is a non-empty, convex, and compact subset. 2) $u_i(b_i; b_{-i})$ is quasi-concave and continuous in $\Omega_i^{RU}$. The optimal solution is $\boldsymbol{d}$, and the optimization problem satisfies the Slater condition [37]. A unique Lagrange multiplier $\mu(p)$ exists that satisfies the following Karush–Kuhn–Tucker (KKT) conditions:

$$\hat{v}_i(d_i) \begin{cases} = \mu(p), d_i > 0 \\ \leq \mu(p), d_i = 0 \end{cases}. \quad (22)$$

To satisfy the condition of Lemma 1, $\boldsymbol{b}$ is the demand-side NE.

*Uniqueness:* Suppose $\boldsymbol{b}$ is the NE on the demand side, known from Lemma 1:

$$\frac{1}{p_i} v_i'(d_i)(1 - \frac{d_i}{P}) \begin{cases} = \mu, d_i > 0 \\ \leq \mu, d_i = 0 \end{cases}, \quad (23)$$

where $\boldsymbol{d}$ is the corresponding energy allocation plan and the KKT condition of the demand-side optimization problem. Using $\mu = \mu(p)$ and $\boldsymbol{d} = d(p)$, shows that $\boldsymbol{b}$ is unique.

*B. Supply Side Game Model*

From (8), the offers of the other ESPs affect the utility function of ESP $j$, and the game on the supply-side is as follows:

1) *Players:* ESPs in set $\Omega_j$.

2) *Strategies:* According to the price weight $q$ received, each ESP determines its supply and offers to minimize energy cost. The strategy set of ESP $j$ is

$$\Omega_j^{ESP} = \{(a_j(s_j), a_{-j}(s_{-j})) | s \in S\}. \quad (24)$$

3) *Payoffs*: $u_j(a_j; a_{-j})$ is the payoff of ESP $j$, as in (8).

**Definition 2:** When each ESP satisfies $u_j(a_j; a_{-j}) \geq u_j(a_j'; a_{-j}), \forall a_j' \geq 0$, offer $\boldsymbol{a}$ is the NE of the game $(J, q)$.

**Lemma 2:** For the supply-side game $(J, q), J > 2$, offer $\boldsymbol{a}$ is the NE if and only if the ESP has

$$\frac{1}{q_j} c_j'(s_j)(1 + \frac{s_j}{(J-2)\sum_{i \in \Omega_i} d_i}) \begin{cases} = v, P > s_j > 0 \\ \leq v, s_j = \sum_{i \in \Omega_i} d_i \\ \geq v, s_j = 0 \end{cases}. \quad (25)$$

**Theorem 2:** On the supply side $(J, q)$, when $J>2$ has a unique NE $\boldsymbol{a}$, for any NE $\boldsymbol{a}$, the supply strategy vector $\boldsymbol{s}$ defined by $s_j = S(\omega(\boldsymbol{a}); a_j)$ is the only solution to (4), and is also the only solution of variational inequality $VI\,(S, N)$, and $N\,(s, q)$ is defined as follows:

$$\hat{c}_j(s_j, q_j) \triangleq \int_0^{s_j} \frac{c_j'(z)}{q_j}(1 + \frac{z}{(J-2)\sum_{i \in \Omega_i} d_i}) dz \quad (26a)$$

$$N_j(s,q) \triangleq -\frac{\partial \hat{c}_j(s_j,q_j)}{\partial d} = -\frac{c'_j(s_j)}{q_j}(1+\frac{s_j}{(J-2)\sum_{i\in\Omega_i} d_i}). \quad (26b)$$

**Proof of the supply-side game:** can be found in [38].

**Theorem 3:** The market reaches the balance of supply and demand if and only if theories 1 and 2 hold, and there exists a unique equilibrium point.

**Proof of the Theorem 3:** can be found in [38].

## IV. SOLUTION METHODOLOGY

In the previous section, a DADP mechanism was devised for power-heating trading to maximize SW.

### A. Performance of the mechanism

A rational market mechanism was engaged in encompassing individual rationality, budget balance, effectiveness, and truthfulness.

*Truthfulness*: The truthfulness of the mechanism was proved by the ability of demand-side and supply-side players to achieve the NE. Under NE, the gain from changing strategy on either side is lower than the gain under NE.

*Effectiveness*: The price weight determined by the ETC is vital for the efficiency of the mechanism. Discriminatory price control using the price weight to maximize market efficiency is adopted in Section IV-B.

*Individual rationality*: Assumptions in Section II, we have given that all players in the market are rational individuals. This is why each player is reluctant to disclose their *private information* to their rivals, except for sharing *public information* such as quotes. However, during the bidding process, the ETC acts as the trusted auctioneer and accesses the information about the offerors to incentivize the players to achieve the maximum overall social welfare. To resolve the conflict between privacy preservation of players and the information requirements of auctioneers, we propose a mechanism for distributed bidding, where each participant can be considered as an independent information node, and the ETC acts as an information center node with each player node having an independent information transmission path that is not known by other adversaries. In market bidding, the process of traders formulating the optimal strategies can be decomposed into the ETC behavior and players behavior, and thus the ADMM algorithm pertains to implementing distributed bidding in Section IV-C.

*Privacy preservation and proof*: In our mechanism, each player only needs to provide its *public information* quotation to the ETC, which is the shared information between the ETC and the player, without revealing any *private information*. ETC can estimate *semi-public information*: the total demand and supply, while issuing *private information* individual price weights which are invisible to the rivals to each player and exchange semi-public information between demand-side and supply-side. No player can infer private information of other players based on their known information set.

We assume that, in the demand side, the set of information that ETC can have access to is $\aleph_D^E = (\mu_i^{nk}, b^{nk}, z_i^{nk}, d_i^{nk}, p_i^{n+1})$, for LA $i \in \Omega_i = \{1,...,I\}$, the known information set is $\aleph_i^D = (\mu_i^{nk}, b^{nk}, z_i^{nk}, d_i^{nk}, p_i^{n+1}, \sum_{j\in\Omega_j} s_j)$; in the supply side, for ETC, the known information set is $\aleph_S^E = (\omega^{nk}, a^{nk}, x^{nk}, s^{nk}, q^{n+1})$; for ESP $j \in \Omega_j = \{1,...,J\}$; the known information set is $\aleph_j^S = (\sum_{i\in\Omega_i} d_i, \omega_j^{nk}, a^{nk}, x_j^{nk}, s_j^{nk}, q_j^{n+1})$. Then, the mathematical definition of privacy preservation proposed in this paper is as follows.

**Definition 3:** For any player $i \in \Omega_i = \{1,...,I\}$, given its information set $\aleph_i^D$, it is not possible to estimate a unique and correct $v_{-i}$ value function. And for any player $j \in \Omega_j = \{1,...,J\}$, it is not possible for itself to estimate a unique and correct cost function $c_{-j}$ based on $\aleph_j^S$.

The proof of privacy preservation of this mechanism is shown in Section IV-F.

*Budget balance*: The budget balance is achieved via the ETC facilitating the achievement of a supply and demand consensus. The ATC algorithm [39] is well suited to the multisystem optimization problem and more flexible than the ADMM algorithm. In this regard, the ATC algorithm is applied to maintain the balance of supply and demand in Section IV-D. The implementation process of DADP mechanism is shown in Fig. 2.

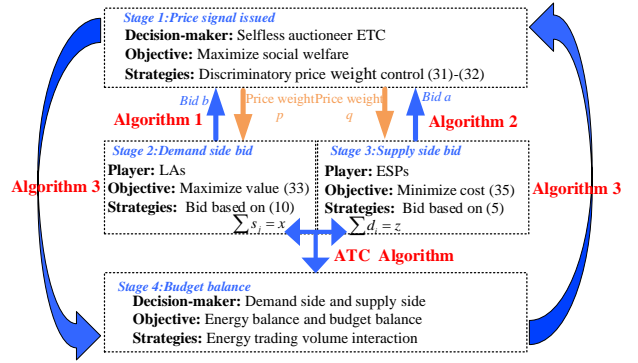

Fig. 2. The implementation process of the DADP mechanism.

### B. Discriminatory Price Weight Control

**Theorem 3:** When the strategy of the player satisfies Theorems 1 and 2, $d$ and $s$ are the NE, and the weights $p$, $q$ can maximize SW if and only if the following equation holds [40]:

$$\frac{p_i}{p_{-i}} = \frac{\sum_{j\in\Omega_j} s_j - d_i(p)}{\sum_{j\in\Omega_j} s_j - d_{-i}(p)}, i,-i \in \Omega_{i,-i\neq i} \quad (27)$$

$$\frac{q_j}{q_{-j}} = \frac{(J-2)\sum_{i\in\Omega_i} d_i + s_j(q)}{(J-2)\sum_{i\in\Omega_i} d_i + s_{-j}(q)}, j,-j \in \Omega_{j,-j\neq j}. \quad (28)$$

Equations (27) and (28) separate the price weight under the NE from each player's strategy, enabling the ETC to converge to the maximum SW without the privacy information of players. When $p^*$, $q^*$ represent the NE point, we can obtain from Theorem 3:

$$\frac{p_i^*}{\sum_{-i\in\Omega_i,i\neq -i} p_{-i}} = \frac{\sum_{j\in\Omega_j} s_j - d_i^*}{(I-1)\sum_{j\in\Omega_j} s_j}, i,-i \in \Omega_{i,-i\neq i} \quad (29)$$

$$\frac{q_j^*}{\sum_{-j\in\Omega_j,j\neq -j} q_{-j}} = \frac{(J-2)\sum_{i\in\Omega_i} d_i + s_j^*}{(J-1)^2 \sum_{i\in\Omega_i} d_i}, j,-j \in \Omega_{j,-j\neq j} \quad (30)$$

Based on the NE solution described by Theorem 3, the price weight is updated as follows:

$$p^{k+1} = p^k + \delta\left(\frac{\sum_{j\in\Omega_j} s_j - d^k}{I-1} - \frac{p^k \sum_{j\in\Omega_j} s_j}{\|p^k\|}\right) \quad (31)$$

$$q^{k+1} = q^k + \delta\left(\frac{(J-2)\sum_{i\in\Omega_i} d_i + s^k}{(J-1)^2} - \frac{q^k \sum_{i\in\Omega_i} d_i}{\|q^k\|}\right) \quad (32)$$

In [41], the researchers proved that when the iterative step size is selected appropriately, the price weight converges to maximize SW.

*Remark 3:* For all players in the market, including LAs and ESPs, price weights are personal and private information, as they adjust their demand and supply based on the weights and then give new offers. If price weights were fully public, players could deduce rival offers based on the available information, thus undermining the fairness of market transactions. Therefore, price weight information is only available to ETC and the players themselves, with each participant receiving a unique price weight. This requires that the information must be interacted with in a fully distributed manner.

### C. ADMM-based Distributed Algorithm

The ADMM-based distributed bidding algorithm (ADBA) was designed considering privacy preservation. The ETC affects the bid strategies of traders through the price weight, and players shape the game strategies based on the price weight and their preferences. The ETC and players alternate actions to reach the NE. Therefore, the auxiliary vector $x \in R^M, z \in R^M$ is introduced, which is the estimate of trading energy $\sum_{i\in\Omega_i} d_i, \sum_{j\in\Omega_j} s_j$ by the ETC.

1) Demand-side bidding algorithm

Demand-side optimization can be translated into the follows:

$$\max_{d,z} \sum_{i\in\Omega_i} v_i(d_i, p_i) - g(z) \quad (33a)$$

$$st. 0 \leq d_i \leq \sum_{j\in\Omega_j} s_j, \forall i \quad (33b)$$

$$d - z = 0, \quad (33c)$$

where $g(z)$ satisfies $g(z) = 0$ when $\|z\| = \sum_{j\in\Omega_j} s_j$; otherwise, $g(z) = +\infty$. The economic meaning of the Lagrange multiplier $\mu$ is the shadow price [42]. The augmented Lagrange of (33) is

$$L_{\rho i} = \sum_{i\in\Omega_i} \hat{v}_i(d_i, p_i) - g(z) - \mu^T(d-z) - \frac{1}{2}\rho\|d-z\|_2^2. \quad (34)$$

Algorithm 1 is used to solve (33) by maximizing $L_{\rho i}$.

| **Algorithm 1**: ADBA on the demand side |
|---|
| 1: **Repeat** |
| 2: ETC sends market signals $(z_i^k, \mu_i^k)$ to each LA. |
| 3: LAs: $d_i^k = \underset{0\leq d\leq \sum_{j\in J} s_j}{argmax}\ \hat{v}_i(d) - \mu_i^k d_i - (\frac{1}{2}\rho)(d_i - z_i^k)^2$ |
| 4: LAs submit bid: $b_i^k = \mu_i^k d_i^k$ |
| 5: ETC update: $z^{k+1} = argmax L_{\rho i}(d^k, z, \mu^k)$ <br> $\mu^{k+1} = \mu^k + \rho(d^k - z^{k+1})$ |
| 6: **Until** $\frac{\|d^k - z^{k+1}\|_1}{\sum_{j\in J} s_j} < \epsilon_{pri}; \frac{\rho\|z^k - z^{k+1}\|_1}{\sum_{j\in J} s_j} < \epsilon_{dual}$ |

2) Supply-side bidding algorithm

Supply-side optimization can be translated as follows:

$$\max_{s,x} \sum_{j\in\Omega_j} g(x) - c_j(s_j, q_j) \quad (35a)$$

$$st. 0 \leq s_j \leq \sum_{i\in\Omega_i} d_i, \forall j \quad (35b)$$

$$x - s = 0, \quad (35c)$$

where $g(x)$ satisfies $g(x) = 0$ when $\|x\| = \sum_{i\in\Omega_i} d_i$; otherwise, $g(x) = +\infty$. The augmented Lagrange of (35) is

$$L_{\rho j} = \sum_{j\in\Omega_j} \hat{c}_j(s_j, q_j) + g(x) + \omega^T(s-x) - \frac{1}{2}\rho\|s-x\|_2^2. \quad (36)$$

The economic meaning of the Lagrange multiplier $\omega$ is the shadow price. Algorithm 2 is used to solve (35) by maximizing $L_{\rho j}$.

| **Algorithm 2**: ADBA on the supply side |
|---|
| 1: **Repeat** |
| 2: ETC sends market signals $(x_j^k, \omega_j^k)$ to each ESP. |
| 3: ESPs: $s_j^k = \underset{0\leq s\leq \sum_{i\in I} d_i}{argmax}\ -\hat{c}_j(s) + \omega_j^k s_j - (\frac{1}{2}\rho)(s_j - \omega_j^k)^2$ |
| 4: ESPs submit offer: $a_j^k = \omega_j^k(\sum_{i\in I} d_i - s_j^k)$ |
| 5: ETC update: $x^{k+1} = argmax L_{\rho j}(s^k, x, \omega^k)$ <br> $\omega^{k+1} = \omega^k + \rho(s^k - x^{k+1})$ |
| 6: **Until** $\frac{\|s^k - x^{k+1}\|_1}{\sum_{i\in I} d_i} < \epsilon_{pri}; \frac{\rho\|x^k - x^{k+1}\|_1}{\sum_{i\in I} d_i} < \epsilon_{dual}$ |

*Remark 4:* From Algorithm 1 and Algorithm 2, the demand-side and supply-side market bidding problems are decomposed into sub-problems for different market participants. On the demand side, the ETC has access to estimated total demand (*semipublic information*) and the LAs' quotes. Each LA can obtain its market signals (*privacy information*) from the ETC, which are not known by the rivals. The supply side is similar to the demand side in that the information is updated in a distributed manner, which perfectly satisfies the need for privacy protection of rational entities. Furthermore, the ADMM algorithm has been proven to work well for distributed convex optimization [42].

### D. ATC Algorithm to Balance Supply and Demand

The ATC algorithm decomposes a system into subsystems, and each subsystem introduces a penalty function to make the coupled variables converge. Through a distributed iterative process, the shared variables of the subsystems' interaction are reached consistently. These characteristics are suitable for enabling the demand side and supply side to achieve an energy balance. The ETC estimates the optimal total supply $\sum_{j\in\Omega_j} \hat{s}_j$ and passes it to the demand side. The LAs shape $b_i$ and $\sum_{i\in\Omega_i} \hat{d}_i$ considering the balance of supply and demand. The two parts are iterated alternately until the convergence conditions are met:

$$\left|\sum_{j\in\Omega_j} s_j^m - \sum_{i\in\Omega_i} d_i^m\right| \leq \varepsilon_1 \quad (37)$$

$$\left|\frac{(\sum_{i\in\Omega_i} v_i^m - \sum_{j\in\Omega_j} c_j^m) - (\sum_{i\in\Omega_i} v_i^{m-1} - \sum_{j\in\Omega_j} c_j^{m-1})}{\sum_{i\in\Omega_i} v_i^m - \sum_{j\in\Omega_j} c_j^m}\right| \leq \varepsilon_2. \quad (38)$$

If (37) and (38) cannot be satisfied simultaneously, the Lagrange multiplier is updated by

$$\chi^m = \chi^{m-1} + 2(\gamma_j^{k-1})^2 (\sum_{j \in \Omega_j} s_j^{m-1} - \sum_{i \in \Omega_i} d_i^{m-1}) \quad (39)$$

$$\gamma_j^m = \beta \gamma_j^{m-1}. \quad (40)$$

The value of β is (2,3), and the initial values of the multipliers $\chi$ and $\gamma$ are generally small constants.

*Remark 5:* In the process of reaching a balance between supply and demand, there is no direct and accurate demand (supply) information interaction between LAs (ESPs) and ETC. The total supply and demand (semipublic information) can only be estimated and transmitted through the ETC, which well protects the private information of players. Fully distributed algorithms like ATC reach the optimal solution through multiple iterations while protecting the private information of independent individuals, as has been demonstrated in the Ref [43].

*E. Distributed Implementation of DADP*

Algorithm 3 summarizes the distributed implementation of DADP which combines price control and ADBA.

| **Algorithm 3**: Distributed double auction algorithm |
|---|
| 1:   **Initialization:** p, ρ, $z^1, \mu^1$, q, $x^1, \omega^1, \chi^1, \gamma^1, s^1$ |
| 2:   ETC advertises parameters to the players |
| 3:   $q^n = q^n/\|q^n\|$; $p^n = p^n/\|p^n\|$ |
| 4:   ETC sends price weight $p_i^n/q_j^n$ to LA $i$/ESP $j$. |
| 5:   k=1; n=1; m=1 |
| 6:   **Repeat** |
| 7:   Do **Algorithm 1** |
| 8:   ETC update: $p^n$: $p^{n+1} = p^{n+1}/\|p^{n+1}\|$ |
| 9:   if $\|p^{n+1} - p^n\|_1 < \epsilon$ output $d_i^m$; otherwise, n=n+1 |
| 10:  ETC sends estimated total demand $z$ to the supply side. |
| 11:  Do **Algorithm 2** |
| 12:  ETC update $q^n$: $q^{n+1} = q^{n+1}/\|q^{n+1}\|$ |
| 13:  if $\|q^{n+1} - q^n\|_1 < \epsilon$ output $s_j^m$; otherwise, n=n+1 |
| 14:  m=m+1, ETC update: $\gamma^m, \chi^m$ |
| 15:  **Until** (37–38) satisfied, output $a_j, b_i, s_j, d_i, q_j, p_i$ |
| 16:  **End** |

The existence and uniqueness of the Nash equilibrium on the demand and supply sides are verified in Theorem 1 and Theorem 2. Accordingly, the demand-side and supply-side Nash equilibria are realized by Algorithm 1 and Algorithm 2. Then, the balance of supply and demand is realized by the ATC algorithm, while Theorem 3 guarantees the existence and uniqueness of the global equilibrium solution. Although Ref. [43] proves that the ATC algorithm can converge to the optimal solution by multiple iterations, the algorithm is more stringent in terms of parameters and does not converge to the exact optimal solution sometimes, in addition to which there are usually some errors and fluctuations in the iterative process. Thus, the theoretical proof for the convergence of this algorithm is still an open problem. Fortunately, the simulation results demonstrated that it has good convergence. Introducing the normalization of the price weight vector in step 3 of Algorithm 3 could improve the numerical stability of the algorithm [26].

*F. Privacy proof*

Here we take the supply side as an example to prove the privacy of ESPs.

Given ESP $j$ the known information set $\aleph_j^S$, for the sake of contradiction, assuming that ESP $j$ can deduce the cost function $c_{-j}$ of ESP $-j$. In this case, the known information set is:

$$\aleph_j^s = (\omega_j^{nk}, a_j^{nk}, x_j^{nk}, s_j^{nk}, q_j^{n+1}, \sum_{i \in \Omega_i} d_i^n), \forall j, -j \in \Omega_{j,-j \neq j} \quad (41)$$

Firstly, to obtain the supply of player ESP $-j$, the ESP $j$ needs to solve the following problem.

$$s_{-j}^{nk} = \arg\max_{0 \le s \le \sum_{i \in I} d_i} -\hat{c}_{-j}(s_{-j}^{nk}) + \omega_{-j}^{nk} s_{-j}^{nk} - (\frac{1}{2}\rho)(s_{-j}^{nk} - \omega_{-j}^{nk})^2 \quad (42)$$

Apparently, part of the variables contained in eq. (42) are elements of the set $\aleph_{-j}$. Comparing the set $\aleph_j^S$ with $\aleph_{-j}$, we have

$$\aleph_{-j} \cap \aleph_j^S = (c_{-j}, \sum_{i \in \Omega_i} d_i^n), \forall j, -j \in \Omega_{j,-j \neq j} \quad (43)$$

Then, the unknown variables for ESP $j$, i.e., the complement of set $\aleph_j^S$ in $\aleph_{-j}$, are as follows

$$\mathbb{C}_{\aleph_{-j}} \aleph_j^S = (\omega_{-j}^{nk}, s_{-j}^{nk}), \forall j, -j \in \Omega_{j,-j \neq j} \quad (44)$$

Therefore, ESP $j$ has to know the information set $\mathbb{C}_{\aleph_{-j}} \aleph_j^S$ to deduce privacy information of player ESP $-j$. However, ESP $j$ has no access to the information set $\mathbb{C}_{\aleph_{-j}} \aleph_j^S$.

Furthermore, ESP $j$ has the public information $a_{-j}^{nk}$, i.e.,

$$a_{-j}^{nk} = \omega_{-j}^{nk} (\sum_{i \in \Omega_i} d_i^n - s_{-j}^{nk}), \forall j, -j \in \Omega_{j,-j \neq j} \quad (45)$$

Based on this, $\omega_{-j}^{nk}$ can be represented by $s_{-j}^{nk}$, as follows

$$\omega_{-j}^{nk} = a_{-j}^{nk} / (\sum_{i \in \Omega_i} d_i^n - s_{-j}^{nk}), \forall j, -j \in \Omega_{j,-j \neq j} \quad (46)$$

By substituting (46) into (42), we have

$$s_{-j}^{nk} = \arg\max_{0 \le s \le \sum_{i \in I} d_i} -\hat{c}_{-j}(s_{-j}^{nk}) + \frac{a_{-j}^{nk}}{\sum_{i \in \Omega_i} d_i^n - s_{-j}^{nk}} s_{-j}^{nk} \\ -(\frac{1}{2}\rho)(s_{-j}^{nk} - \frac{a_{-j}^{nk}}{\sum_{i \in \Omega_i} d_i^n - s_{-j}^{nk}})^2 \quad (47)$$

Therefore, if ESP $j$ has the information $s_{-j}^{nk}$ or $\omega_{-j}^{nk}$, and the number of iterations $k$ is large enough, the player $j$ is able to obtain a unique cost function curve of ESP $-j$. However, in (48), for ESP $j$, there is only one equation, but two variables $c_{-j}$ and $s_{-j}^{nk}$. Thus, ESP $j$ is not able to derive $c_{-j}$ based on the known information set $\aleph_j^S$.

This completes the proof.

Similarly, we can prove any LA $i$ in the demand side cannot deduce any private information from their known information sets.

V. PERFORMANCE EVALUATION

To demonstrate the effectiveness of DADP and its algorithm, a hypothetical electric-heating market of a city-level distribution system consisting of five ESPs and five LAs was used to perform the simulation. During the trading process, the players dynamically join or leave the market. The entities in the transaction in different periods are shown in Table I. The specific parameters of the market players can be found in [38]. The simulations were performed on a 64-bit laptop with 4GB-RAM and i5-4210M CPU-2.6GHz using CLPEX and YALMIP.

TABLE I
PERIOD OF TRADING ENTITIES

| Scene | The participators of the transaction | |
|---|---|---|
| S1 | LA1, LA2 | ESP1, ESP2, ESP3 |
| S2 | LA1, LA2, LA3 | ESP1, ESP2, ESP3 |
| S3 | LA1, LA2, LA3, LA4 | ESP1, ESP2, ESP3 |
| S4 | LA1, LA2, LA3, LA4, LA5 | ESP1, ESP2, ESP3 |
| S5 | LA1, LA2, LA3, LA4, LA5 | ESP1, ESP2, ESP3, ESP4 |
| S6 | LA1- LA5 | ESP1-ESP5 |
| S7 | LA2, LA3, LA4, LA5 | ESP1-ESP5 |
| S8 | LA4, LA3, LA5 | ESP2, ESP3, ESP4, ESP5 |
| S9 | LA3, LA5 | ESP3, ESP5 |

*A. Demand Side Analysis*

Fig. 3 shows the NE dynamic changes of the LAs in the power market.

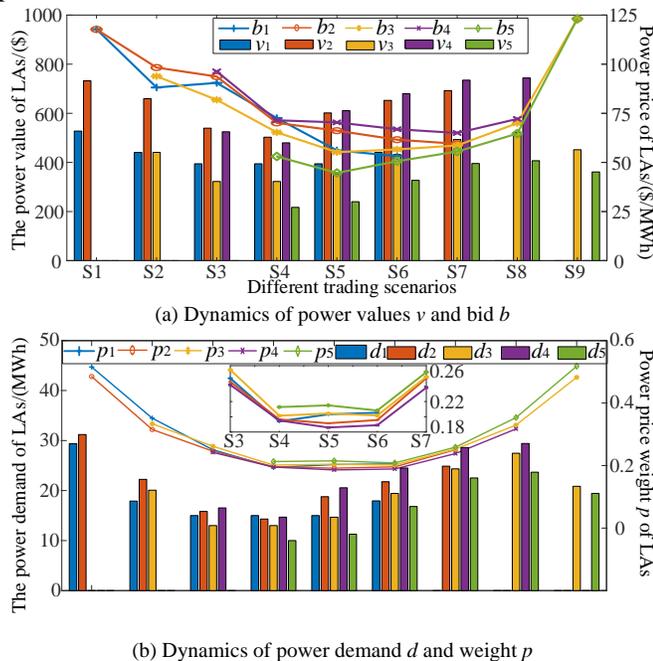

(b) Dynamics of power demand $d$ and weight $p$
Fig. 3. Power demand-side dynamics under different scenarios.

As the number of LAs in S1–S4 increases, the market share competition on the demand side increases, the LAs' bid decreases; similarly, the market share of each LA also reduces. By contrast, in S6–S9, as the number of LAs participating in the competition decreases, to obtain market share, the demand-side bids increase.

Fig. 4 shows the NE dynamic changes of the LAs in the heat market. The heating market and power market have different characteristics because user thermal comfort is considered. Although the market entities change, heat demand barely shifts. Additionally, from S1–S3, the heating price increases because the demand of the original participants is almost at a constant level, and the increase in market competitors inevitably causes an uptick in bidding to meet the lowest heat demand. Interestingly, the heating price dropped from S3 to S4 with the entry of LA5 into the market. The reasons resulting in price reductions mainly are both the high heat demand from LA5 and the highest marginal value of heat use. Consequently, the biggest market share goes to LA5 with the highest bids, which leads to a moderation in a lower price of competitors.

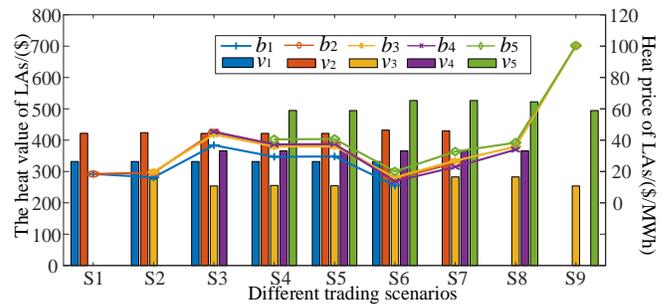

(a) Dynamics of heat values $v$ and bid $b$

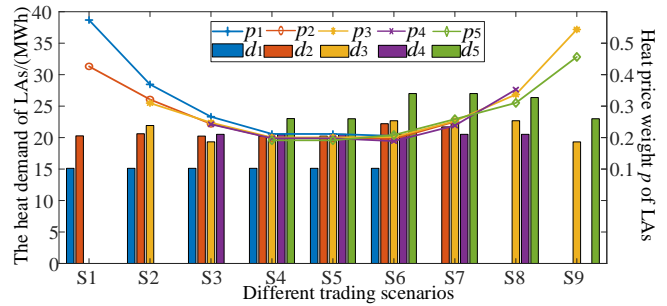

(b) Dynamics of heat demand $d$ and weight $p$
Fig. 4. Heat demand-side dynamics under different scenarios.

As an auctioneer, the ETC aims to maximize the value of energy, prioritizing LAs with high-value functions to enable them to purchase more energy by reducing their corresponding weight. LAs with a high-value function raise bids to occupy more resources; thus, they have a higher price. LAs with the lowest value function almost only meet the minimum energy purchasing limit; hence, more energy is supplied to LAs with a higher value function to increase overall market efficiency.

*B. Supply-Side Analysis*

The results of the supply-side electric market are shown in Fig. 5. The greater the ETC published price weight $q$, the lower the offer of the ESP, and simultaneously the larger the energy supply. As the number of demand-side participants increases, the market demand in S1–S4 increases. Given this context, suppliers have a profit advantage in the market, so they take the opportunity to increase prices to engage in arbitrage. From S5 to S7, as the number of ESPs increases, the supply side becomes keener to compete, and the ESPs have to lower their offers to increase market share. A decrease in the number of market players should reduce the overall offer and supply in S7–S9.

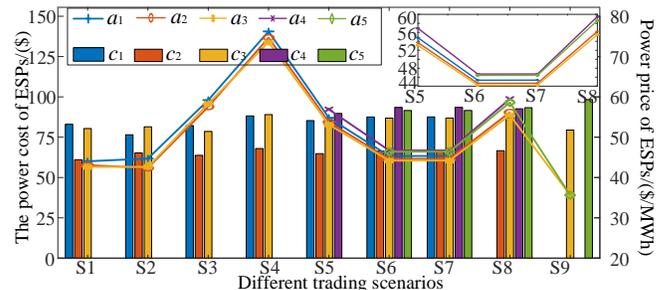

(a) Dynamics of power cost $c$ and offer $a$

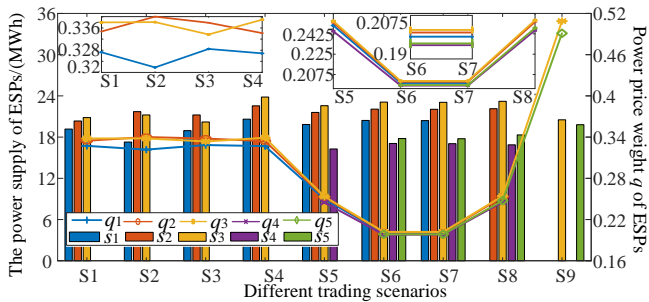
(b) Dynamics of power supply *s* and weight *q*
Fig. 5. Power supply-side dynamics under different scenarios.

However, to meet the electricity demand in S8, except for ESP1 withdrawing from market competition, the power supply of the other suppliers is the same as in S7, but considering the reduction in the number of competitors, each ESP increases its offer to explore more market share. Interestingly, comparing the offer in Fig. 5 (a) with the corresponding supply of ESPs in Fig. 5 (b), the offer of the ESP is always inversely proportional to the supply, this is because the offer of the ESP indicates the level of reluctance to provide energy, measured in US dollars. Moreover, the supply function is introduced in Eq. (5), where the offer is inversely proportional to the supply.

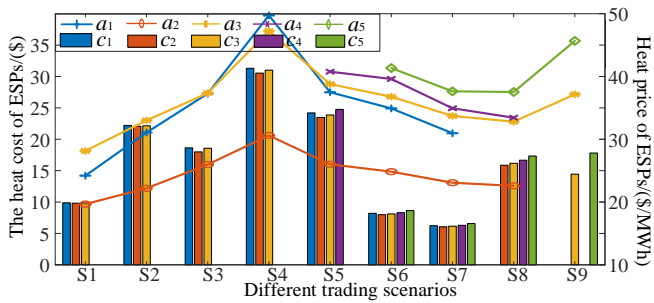
(a) Dynamics of heat cost *c* and offer *a*

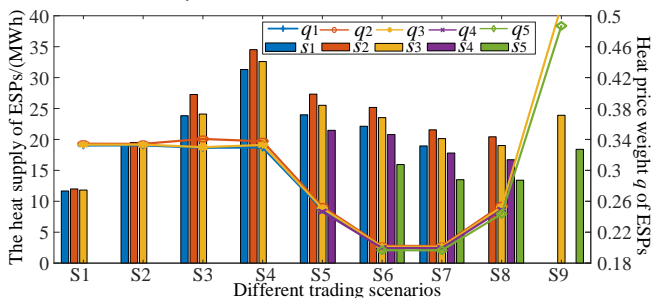
(b) Dynamics of heat supply *s* and weight *q*
Fig. 6. Heat supply-side dynamics under the different scenarios.

The characteristics of the supply-side heat market are similar to the electricity market, except that the heat price curve changes more smoothly and the fluctuations are smaller. To maximize SW, the ETC reduces the total cost as much as possible. Consequently, the ETC gives priority to suppliers with low costs by increasing the corresponding weight and simultaneously actively knocks down the price. New entrants reduce the market share of previous stakeholders.

### C. Analysis of Computational Efficiency

The convergence process of the distributed double-sided auction algorithm is illustrated in Fig. 7.

To demonstrate the convergence performance of the algorithm, the S6 iteration with the most participants in the power market is selected for specific analysis. Fig. 7 shows that the number of iterations on the demand-side and the supply-side does not exceed 10. When the number of iterations is less than four, there is more tremendous shock, but when the number of iterations reaches six, the convergence curve gradually stabilizes.

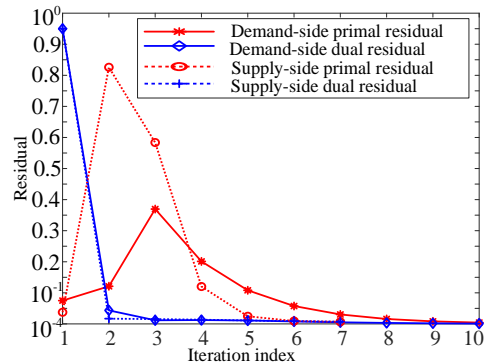
Fig. 7. Convergence process of the distributed double-auction algorithm.

### D. Impact of the Number of Players

The DADP mechanism designed in this paper is obtained by improving and optimizing on the basis of the Kelly mechanism (KEL) in [44]. Therefore, we compared and validated the DADP mechanism with the Kelly mechanism in the process of parameter sensitivity analysis.

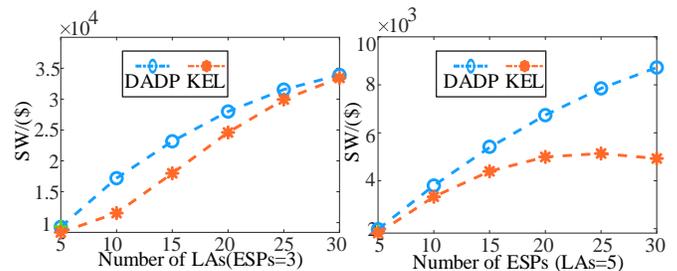
(a) SW varies with the number of players in different mechanisms

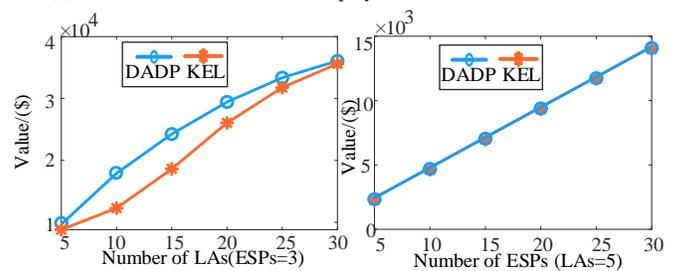
(b) Value varies with the number of players in different mechanisms

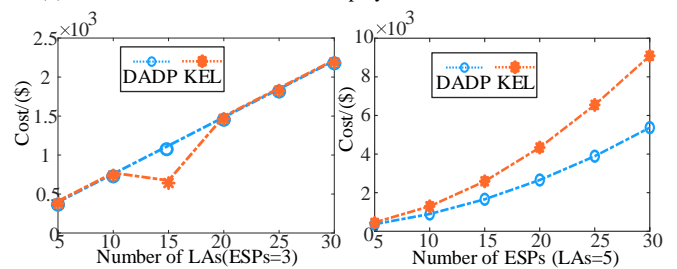
(c) Cost varies with the number of players in different mechanisms
Fig. 8. Comparison of SW for different numbers of players.

The comparison of SW for different numbers of players is shown in Fig. 8. For the demand side, under the scenario with three suppliers, the number of LAs increases from 5 to 30. As the number of LAs increases, although the value of power is almost constant, the cost of DADP is significantly lower than that of KEL. The greater the sales energy, the more cost savings can be achieved using the KEL. Thus, SW constantly improves.

For the supply side, under the scenario with five LAs, the number of ESPs increases from 5 to 30. Interestingly, as the number of ESPs increases, the cost of the three methods barely shifts, but the value of energy changes obviously. Compared with KEL, the energy value of DADP increases first and then decreases. The results show that in scenarios in which the number of suppliers increases, the effect of DADP in terms of improving SW is limited. To summarize, SW improves as the number of players increases, and as the volume of energy transactions increases, the DADP mechanism enhances SW more than the KEL.

*E. Comparative Analysis of Market Mechanisms*

The DADP mechanism proposed in this paper is compared with three other approaches, i.e., the Kelly mechanism, the marginal price clearing mechanism in the power pool [45], and the VCG mechanism [46]. The main characteristics of these classical mechanisms are as follows: 1) The proposed DADP mechanism is improved of Kelly mechanism; 2) The electricity pooling is a mature and centralized form of trading electricity, it creates a unilateral market for power generation where users do not participate in the offer; 3) The VCG mechanism is a widely used distributed trading mechanism without budget balance, although it is capable of maximizing SW while ensuring truthfulness and individual rationality. We compare these mechanisms to verify the effectiveness and superiority of the proposed mechanism. The scenarios in Table I are used to comprehensively analyze SW changes under different mechanisms, including 33 LAs and 32 ESPs.

TABLE II
COMPARATIVE OF MARKET MECHANISMS

| Mechanism | Energy | Cost ($) | Value ($) | SW($) |
|---|---|---|---|---|
| **DADP** | Power | **2612.43** | **16033.88** | **13421.45** |
| | Heat | **1076.55** | **12362.83** | **11286.29** |
| KEL [44] | Power | 2770.29 | 15222.10 | 12451.81 |
| | Heat | 1177.74 | 12253.17 | 11075.43 |
| POOL [45] | Power | 2818.90 | 15796.92 | 12978.02 |
| | Heat | 1081.87 | 12259.84 | 11177.97 |
| VCG [46] | Power | 2612.22 | 16034.06 | 13421.84 |
| | Heat | 1075.94 | 12363.41 | 11287.47 |

The comparison results in Table II show that the market efficiency is lowest under the Kelly mechanism, followed by a lower SW under the traditional power pool model. Ignoring the bias in the calculations, the social welfare under the DADP mechanism and the VCG mechanism are almost equal. Although the market can converge to the state of maximum social welfare under both the VCG and DADP mechanisms, the VCG mechanism cannot autonomously reach budget balance. The advantage of the DADP mechanism proposed in this paper is precisely the ability to maximize social welfare while balancing the budget without the need for third-party monetary subsidies. Comparing the electricity and heat market we can find that the social utility of DADP in the heat market is not as effective as that in the electricity market because user thermal comfort is considered. In summary, the DADP mechanism can improve market efficiency by 10% and 4% compared to the Kelly mechanism and the power pool model, respectively.

## VI. CONCLUSION

In this paper, a DADP mechanism with discriminatory pricing was designed to maximize SW in the power and heating market, in which ESPs submit offers and LAs submit bids to the ETC. As a selfless auctioneer, the ETC uses a discriminatory price weight to incentivize players' strategic bidding, integrating their selfish interests and overall SW. The interactions among players with the same market role are described by a Nash game. An ADMM-based decentralized bidding algorithm was developed to achieve decision-making, considering privacy preservation, whereas the ATC algorithm was used to achieve a balanced amount of trading energy.

Case studies show that the DADP mechanism improved SW by 4%–15% compared with other mechanisms and the algorithm had a good convergence effect. Furthermore, the existence and uniqueness of the NE were also proved in detail. The breakthrough of this paper is the manipulation of the price weight to incentivize selfish players to reach the NE while SW is optimal.

This work focuses on the design of energy trading market mechanisms with privacy-preserving to maximize social welfare. Hence the complex physical constraints, such as energy storage, gas turbines, etc., are not considered in this paper. However, the DADP trading mechanism designed in this paper can well meet the privacy protection needs of players in the market bidding process, and the combination with privacy protection technologies such as blockchain will bring inspiration for the design of novel distributed energy market mechanisms.

Future research can focus on the transparent energy trading mechanisms for prosumers to integrate their selfish interests and overall SW while taking into account complex physical constraints. Privacy protection has always been a primary consideration for market players and an urgent issue to be addressed in the design of future energy market mechanisms.